\documentclass{pasj00}
\usepackage{natbib}
\draft

\begin{document}
\SetRunningHead{Author(s) in page-head}{Running Head}

\title{A Long-period Eccentric Substellar Companion to the Evovled
Intermediate-Mass Star HD\,14067
}

\author{
    Liang \textsc{Wang}\altaffilmark{1}
    Bun'ei \textsc{Sato}\altaffilmark{2}
    Masashi \textsc{Omiya}\altaffilmark{2}
    Hiroki \textsc{Harakawa}\altaffilmark{2}
    Yujuan \textsc{Liu}\altaffilmark{1}
    Nan \textsc{Song}\altaffilmark{1,3}
    Wei \textsc{He}\altaffilmark{1,3}
    Xiaoshu \textsc{Wu}\altaffilmark{1,3}
    Hideyuki \textsc{Izumiura}\altaffilmark{4,5}
    Eiji \textsc{Kambe}\altaffilmark{4}
    Yoichi \textsc{Takeda}\altaffilmark{5,6}
    Michitoshi \textsc{Yoshida}\altaffilmark{7}
    Yoichi \textsc{Itoh}\altaffilmark{8}
    Hiroyasu \textsc{Ando}\altaffilmark{6}
    Eiichiro \textsc{Kokubo}\altaffilmark{6}
    Shigeru \textsc{Ida}\altaffilmark{2}
    and
    Gang \textsc{Zhao}\altaffilmark{1}
}
\altaffiltext{1}{
    Key Laboratory of Optical Astronomy,
    National Astronomical Observatories,
    Chinese Academy of Sciences,
    20, Datun Road, Chaoyang District, Beijing 100012, China
}
\email{gzhao@nao.cas.cn}
\altaffiltext{2}{
    Tokyo Institute of Technology,
    2-12-1 Ookayama, Meguro-ku,
    Tokyo 152-8550, Japan
}
\altaffiltext{3}{
    University of Chinese Academy of Sciences,
    19A Yuquan Road, Shijingshan District,
    100049 Beijing, China
}
\altaffiltext{4}{
   Okayama Astrophysical Observatory, National
   Astronomical Observatory of Japan, Kamogata,
   Okayama 719-0232, Japan
}
\altaffiltext{5}{
   The Graduate University for Advanced Studies,
   Shonan Village, Hayama, Kanagawa 240-0193, Japan
}
\altaffiltext{6}{
   National Astronomical Observatory of Japan, 2-21-1 Osawa,
   Mitaka, Tokyo 181-8588, Japan
}
\altaffiltext{7}{
   Hiroshima Astrophysical Science Center, Hiroshima University,
   Higashi-Hiroshima, Hiroshima 739-8526, Japan
}
\altaffiltext{8}{Nishi-Harima Astronomical Observatory, Center for Astronomy,
   University of Hyogo, 407-2, Nishigaichi, Sayo, Hyogo
   679-5313, Japan
}


\maketitle

\begin{abstract}
We report the detection of a substellar companion orbiting an evolved
    intermediate-mass ($M_\star=2.4\,M_\odot$) star HD\,14067 (G9\,III) using
    precise Doppler technique.
Radial velocities of this star can be well fitted either by a periodic Keplerian
    variation with a decreasing linear velocity trend ($P=1455$ days, $K_1=92.2$
    m\,s$^{-1}$, $e=0.533$, and $\dot{\gamma}=-22.4$ m\,s$^{-1}$\,yr$^{-1}$) or
    a single Keplerian orbit without linear trend ($P=2850$ days, $K_1=100.1$
    m\,s$^{-1}$, and $e=0.697$).
The minimum mass ($m_2\sin{i}=7.8\,M_{\rm J}$ for the model with a linear trend,
    or $m_2\sin{i}=9.0\,M_{\rm J}$ for the model without a linear trend)
    suggests a long-period giant planet around an evolved intermediate-mass star.
The eccentricity of the orbit is among the highest known for planets ever
    detected around evolved stars.
\end{abstract}

\section{Introduction}
Since the first giant planet orbiting evolved star was discovered by \citet{
    Frink2002}, progress has been made on the detection and theoretical
    understanding of planet around stars more massive than our Sun in the past
    decade.
Over 90 substellar companions with minimum masses ranging from 0.6 to 40
    $M_{\rm J}$ around giants and subgiants have been detected by precise radial
    velocity technique (e.g. \citealt{Sato2003, Setiawan2003, Hatzes2005,
    Johnson2007, Sato2008a, Dollinger2009, Liu2009, Wittenmyer2011, Johnson2011,
    Omiya2012, Gettel2012, Lee2013}), including those in open clusters (e.g.
    \citealt{Sato2007, Lovis2007, Brucalassi2014}), in multiple-planet systems
    (e.g. \citealt{Niedzielski2009, Sato2013a}), and with masses $\gtrsim13\,
    M_{\rm J}$ and thus lie in the brown dwarf regime (e.g. \citealt{Omiya2009,
    Sato2010, Wang2012}).

Although the number of such companions around GK giants is still insufficient to
    make a statistical study of their physical properties, they are of intense
    interest because their host stars are the slowly rotating counterparts of
    intermediate-mass ($1.5<M_\star/M_\odot<5$) B-A dwarfs that have evolved off
    the main sequence, giving us a chance to study planets orbiting stars with
    masses larger than those of FGK dwarfs.
Recent studies have revealed some distinct differences from the FGK main
    sequence stars.
For instance, nearly all detected planets around $M_\star>1.5\,M_\odot$ stars
    have semi-major axes $\gtrsim0.6$\,AU, with only a few exceptions including
    HD\,102956\,b \citep{Johnson2010a}, WASP-33\,b \citep{CollierCameron2010},
    and Kepler-13\,Ab \citep{Szabo2011}.
Such paucity can be attributed to the engulfment by the host stars as they
    evolved off the main-sequence \citep{Sato2008b, Nordhaus2010}, or the
    primordial deficiency of short-period planets during their formations (e.g.
    \citealt{Currie2009, Kretke2009}).
On the other hand, most planets ever detected around intermediate-mass stars
    have eccentricities bellow 0.4.
It is natural because most of their hosts are in the post-RGB (core helium
    burning) phase, and the planetary orbits could have been tidally
    circularized due to the increasing radii of the host stars as it ascends the
    red giant branch.
However, some planets with high eccentricities ($e>0.6$) have been discovered
    (e.g. \citealt{Sato2013b, Moutou2011, Niedzielski2009}), implying the
    existence of planet-planet scattering scenario (e.g. \citealt{Ford2008}) or
    perturbators (e.g. \citealt{Takeda2005}).

In this paper, we report on the detection of a new substellar companion around
    an intermediate-mass giant HD\,14067 from our planet search program using
    the Subaru 8.2m telescope, the OAO 1.88m telescope and the Xinglong 2.16m
    telescope.
Observations are described in section 2 and stellar properties are summarized in
    section 3.
The analyses of radial velocities and orbital solutions are given in section 4.
In section 5 we give the conclusion.

\section{Observations and Radial-Velocity Analysis}

We have been conducting a precise radial velocity survey for about 300 G-K
    giants at Okayama Astrophysical Observatory (OAO) Japan since 2001.
To extend this planet search program, we established an international network
    among Japanese, Korean, and Chinese researchers using three 2m class
    telescopes in 2005 (East-Asian Planet Search Network; \citealt{
    Izumiura2005}), and started the Subaru planet search program in 2006.
By taking advantage of the large aperture (8.2m) of the Subaru Telescope, the
    planet-hosting candidates among a sample of $\sim$300 giants are quickly
    identified, and the visual magnitudes of $6.5\leq V\leq7.0$ enable them to
    be subsequently followed up by 2m class telescopes.
For the details of the Subaru planet search program, readers are referred to the
    description by \citet{Sato2010}.

We obtained a total of 3 spectra for HD\,14067 in 2007 September, 2008 January
    and August using the High Dispersion Spectrograph (HDS; \citealt{
    Noguchi2002}) equipped with the Subaru Telescope.
An iodine (I$_2$) absorption cell was used to provide a fiducial wavelength
    reference for precise radial velocity measurements \citep{Kambe2002,
    Sato2002}.
We adopted the setup of StdI2b in the first two runs and StdI2a in the third
    one, which covered a wavelength region of 3500-6200 \AA\ and 4900-7600 \AA,
    respectively.
The slit width was set to $0''.6$, giving a resolving power ($\lambda/\Delta
    \lambda$) of 60,000.
The typical signal-to-noise ratio (S/N) was 140--200 pixel$^{-1}$ with an
    exposure time of 30--50s.

After the observations at the Subaru Telescope, we started the follow-up
    observations using the 1.88m telescope with the High Dispersion Echelle
    Spectrograph (HIDES; \citealt{Izumiura1999}) at OAO.
The wavelength region was set to simultaneously cover 3750-7500\,\AA\ using the
    RED cross-disperser with a mosaic of three CCDs.
We set the slit width to 200\,$\mu$m ($0''.76$), giving a resolving power
    ($\lambda/\Delta\lambda$) of 67,000 with 3.3 pixel sampling, and used an
    iodine cell for precise wavelength calibration.
We collected a total of 27 data points of HD\,14067 with HIDES from October,
    2008 to January, 2014.

Since November 2012, we started the follow-up observations of HD\,14067 with the
    High Resolution Spectrograph (HRS) attached at the Cassegrain focus of the
    2.16m telescope at Xinglong Observatory, China.
The fiber-fed spectrograph is the successor of the Coud{\'e} Echelle
    Spectrograph (CES; \citealt{Zhao2001}), giving higher wavelength resolution
    and optical throughput.
The single 4K $\times$ 4K CCD covers a wavelength region of 3700-9200\,\AA.
The slit width was set to 190\,$\mu$m, corresponding to a resolving power
    ($\lambda/\Delta\lambda$) of 45,000 with 3.2 pixel sampling.
An iodine cell was installed before the fiber entrance to obtain the precise
    wavelength reference.

The reduction of the echelle spectra was performed using the IRAF \footnote{
    IRAF is distributed by the National Optical Astronomical Observatory, which
    is operated by the Association of Universities for Research in Astronomy,
    Inc., under cooperative agreement with the National Science Foundation.}
    software package in the standard manner.
The I$_2$-superposed spectra are modeled based on the algorithm given by \citet{
    Sato2002, Sato2012}.
The stellar template used for radial velocity analysis was extracted by
    deconvolving an instrumental profile, which was determined from a spectrum
    of a B-star taken through the I$_2$ cell \citep{Sato2012}.

\section{Stellar Properties}

\begin{figure}
    \begin{center}
    \FigureFile(12cm,){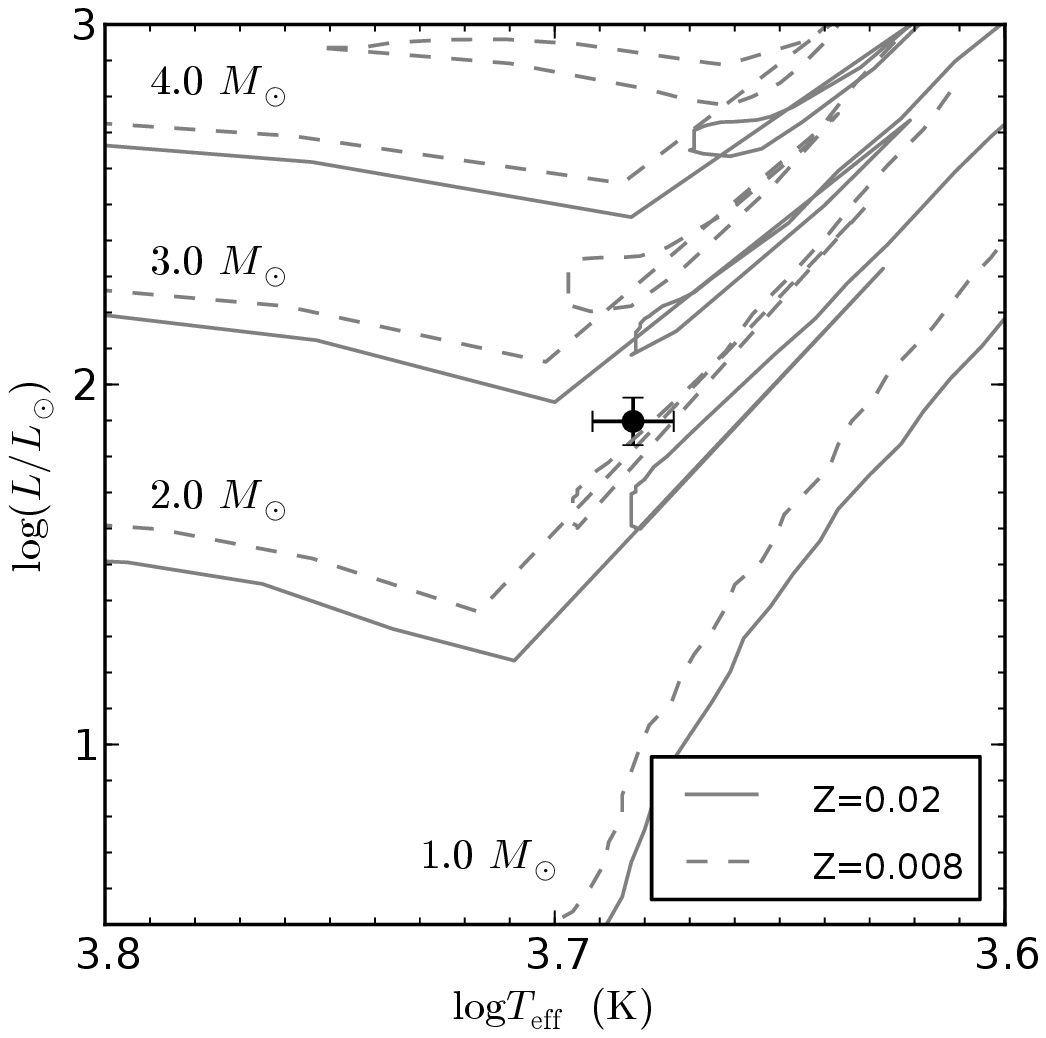}
    \end{center}
    \caption{H-R diagram. The solid cricle represents HD\,14067, with the
        errorbars corresponding to the uncertainties given in table
        \ref{tbl:star}.
        The solid and dashed lines reprensent the evolution tracks from
        \citet{Lejeune2001} for stars of $M=1\sim4\,M_\odot$ with $Z=0.02$
        (solar metallicity) and $Z=0.008$, respectively.
    }\label{fig-hrd}
\end{figure}

\begin{figure}
    \begin{center}
    \FigureFile(12cm,){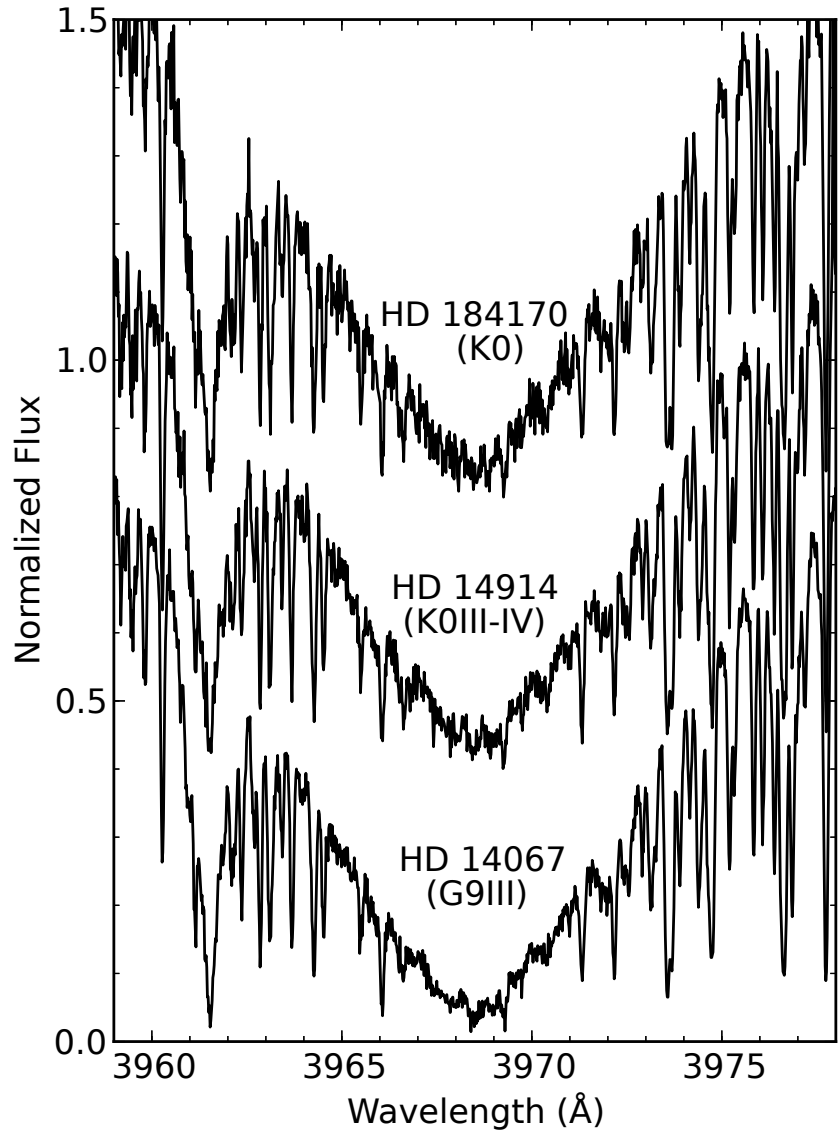}
    \end{center}
    \caption{Spectra in the region of Ca H lines.
        Stars with similar spectral type to HD\,14067 are also shown in this
        figure.
        A vertical offset of 0.4 is added to each spectrum.}\label{fig-CaIIH}
\end{figure}

HD\,14067 (HIP\,10657, BD\,+23\,307, HR\,665, TYC\,1765-1369-1) is listed in the
    {\sc Hipparcos Catalogue} \citep{Perryman1997} as a G9\,III star, with a
    visual magnitude of $V=6.53$ and a color index of $B-V=1.025$.
Its Hipparcos parallax $\pi=6.12\pm0.79$\,mas \citep{vanLeeuwen2007} corresponds
    to a distance of $163.4\pm12.3$ pc and an absolute magnitude of $M_V=0.33$.
The effective temperature $T_{\rm eff}=4815\pm100$ K and bolometric
    correction $BC=-0.329$ were derived from the color index $B-V$ and the
    estimated metallicity using the empirical calibration of \citet{Alonso1999,
    Alonso2001}.
The color excess $E(B-V)$ was calibrated according to the reddening estimation
    given by \citet{Schlegel1998}, and the interstellar extinction was found to
    be $A_V=0.13$.
The surface gravity $\log{g}=2.61\pm0.10$ was determined from the triangular
    parallax given by the new reduction of the {\sc Hipparcos Catalogue}
    \citep{vanLeeuwen2007}.
The iron abundance [Fe/H] was determined from the equivalent widths of $\sim$30
    unblended Fe lines measured from an iodine-free stellar spectrum taken with
    HIDES, and the LTE model atmosphere adopted in this work were interpolated
    from the ODFNEW grid of ATLAS9 \citep{Castelli2004}.
The stellar mass, radius and age were estimated using a Bayesian approach
    similar to that of \citet{daSilva2006}.
We used the Geneva database \citep{Lejeune2001}, which covers the phases from
    the main-sequence to the early asymptotic giant branch (EAGB) stages for
    stars with $2\le M/M_\odot\le5$, to interpolate an extensive grid of stellar
    evolutionary tracks, with $\Delta M=0.05$ within $1.2\le M/M_\odot\le3.6$,
    $\Delta$[Fe/H] = 0.02 within -0.4 $\le$ [Fe/H] $\le$ +0.3, and 500
    interpolated points in each track, spanning the whole evolutionary history.
For each data points, the likelihood functions of $\log{L}$, $T_{\rm eff}$,
    and [Fe/H] were calculated to match the observed values by assuming a
    Gaussian error for each parameter.
To simplify the calculation, we adopted uniform prior probabilities of mass and
    [Fe/H], but weighted the probability of each model with its spanning age
    along its evolutionary track.
The details of the method will be described in an upcoming paper by
    \citet{Wang2014}.
The probability distribution functions (PDFs) of the parameters yield $M=2.4\pm
    0.2\,M_\odot$, $R=12.4\pm1.1\,R_\odot$ and age $=0.69\pm0.20$ Gyr, and the
    probability that the star has passed through the RGB tip and in core helium
    burning phase is $\sim$97\%.
In figure \ref{fig-hrd}, we plotted the HD\,14067 on H-R diagram, together with
    the evolutionary tracks from \citet{Lejeune2001} of stars with different
    masses and metal content.
The macro turbulence velocity was estimated with the empirical relations of
    \citet{Hekker2007} and the stellar rotational velocity $v\sin{i}$ is less
    than 1 km\,s$^{-1}$ according to \citet{deMedeiros1999}.
We also determined the lithium abundance by fitting the line profile of Li\,I
    $\lambda$\,6707.8\,\AA\ doublet using the spectra synthesis method with the
    IDL/SIU package.
The results of $\log{A({\rm Li})}_{\rm LTE}=0.53$ and $\log{A({\rm Li})_{\rm
    NLTE}}=0.73$ suggest HD\,14067 is a Li-depleted giant \citep{Liu2014}.
The derived stellar parameters of HD\,14067 are listed in table \ref{tbl:star}.
Furthermore, the star shows no significant emission in the core of Ca II HK
    lines, as shown in figure \ref{fig-CaIIH}, which suggests that HD\,14067 is
    chromospherically inactive.

\section{Radial Velocities and Orbital Solutions}
\begin{figure}
    \begin{center}
    \FigureFile(16cm,){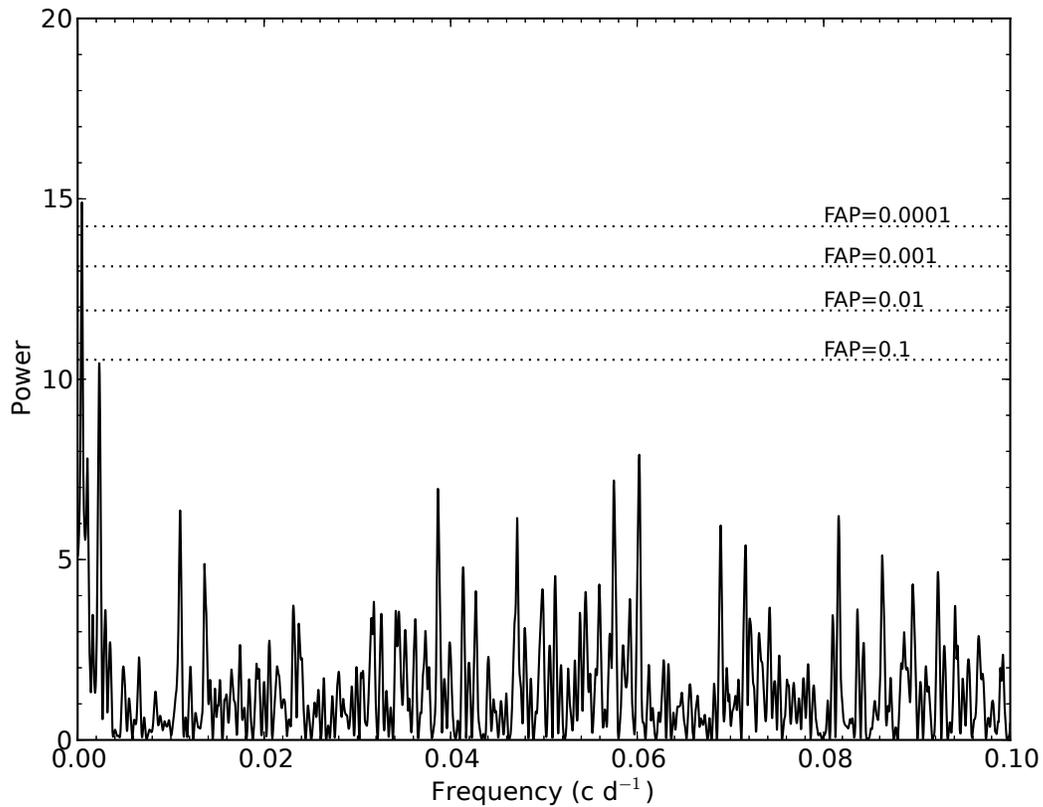}
    \end{center}
    \caption{
        Generalized Lomb-Scargle periodogram of the radial velocities of
        HD\,14067.
        The normalization and FAP (false-alarm probability) were calculated
        according to \citet{Horne1986}.
        There is a significant peak at $f\sim4.9\times10^{-4}$\,c\,d$^{-1}$,
        with the FAP $<1\times10^{-6}$.
        The dash horizontal lines indicate the FAP levels of $10^{-1}$,
        $10^{-2}$, $10^{-3}$, and $10^{-4}$, respectively, from the bottom to
        the top.
    }\label{fig-gls-rv}
\end{figure}

We obtained a total of 52 velocity data points (27 from OAO, 22 from Xinglong,
    and 3 from Subaru) of HD\,14067 over a span of more than 5 years.
The radial velocities are listed in table \ref{tbl-rv} together with their
    estimated uncertainties.
The generalized Lomb-Scargle periodogram \citep{Zechmeister2009} of HD\,14067
    shows a significant peak with FAP (false-alarm probability)
    $<1\times10^{-6}$ near the frequency of $4.9\times10^{-4}$ c\,d$^{-1}$
    (figure \ref{fig-gls-rv}).
We also noticed that the star seems to exhibit a descendent linear velocity
    trend besides the periodic variability.
The significance of the trend depends on how large systematic error is included
    in the first three Subaru data points, but it is difficult to be estimated.
Therefore, we performed the least-squared orbital fitting by a single Keplerian
    with and without a linear trend, simultaneously.
The orbital parameters and the uncertainties were derived using the Bayesian
    Markov Chain Monte Carlo (MCMC) method (e.g., \citealt{Ford2005,Gregory2005,
    Ford2007}), following the analysis in \citet{Sato2013b}.
We took account of velocity offsets of Xinglong and Subaru data relative to OAO
    data, $\Delta$RV$_{\rm Xinglong-OAO}$ and $\Delta$RV$_{\rm Subaru-OAO}$, as
    free parameters in the orbital fitting.
Extra Gaussian noises for each of the three data sets, $s_{\rm OAO}$, $s_{\rm
    Xinglong}$ and $s_{\rm Subaru}$, including intrinsic stellar jitter as well
    as unknown noise source were also incorporated as free parameters.
We generated 5 independent chains having $10^7$ points with acceptance rate of
    about 25\%, the first 10\% of which were discarded, and confirmed each
    parameter was sufficiently converged based on the Gelman-Rubbin statistic
    \citep{Gelman1992}.
We derived the median value of the merged posterior probability distribution
    function (PDF) for each parameter and set 1$\sigma$ uncertainty as the range
    between 15.87\% and 84.13\% of the PDF.

\begin{figure}
    \begin{center}
    \FigureFile(16cm,){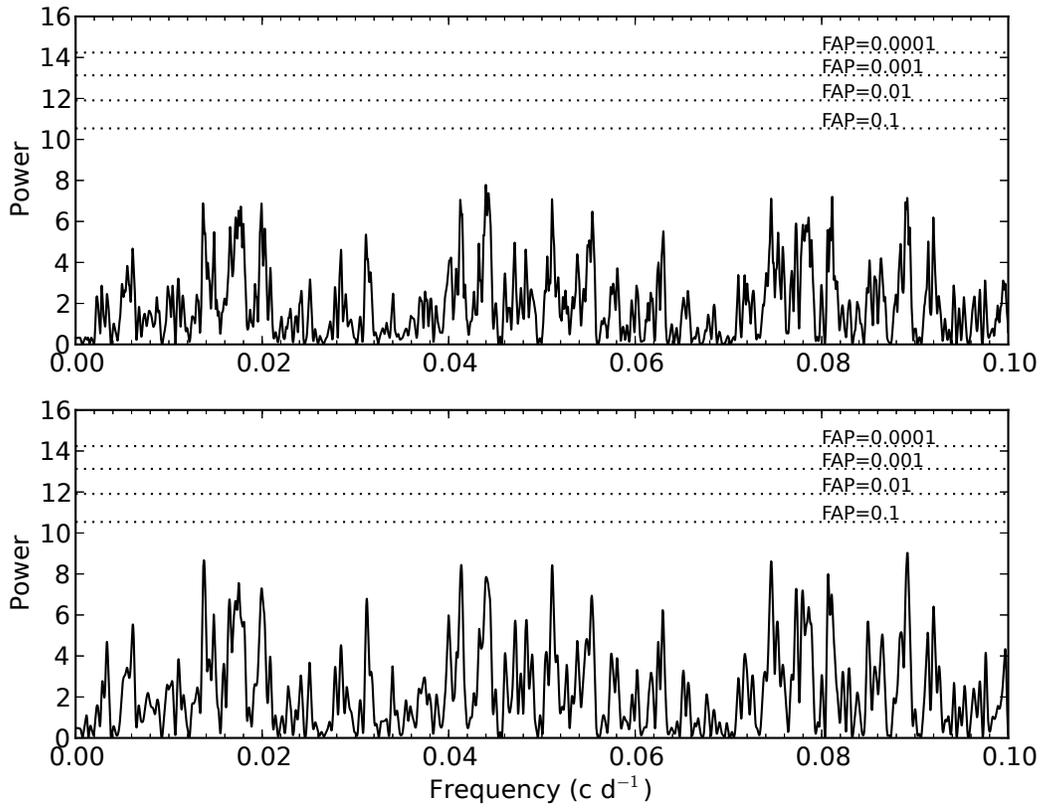}
    \end{center}
    \caption{
        Generalized Lomb-Scargle periodograms of the radial velocity residuals
        after Keplerian orbital fitting with ({\it upper panel}) and without
        ({\it lower panel}) a linear trend.
        Neither of them shows any significant peak with FAP above 0.1.
    }\label{fig-gls-res}
\end{figure}

In figure \ref{fig-rv} we plot the best-fit Keplerian orbit with a linear
    velocity trend, together with the measured data points and their
    uncertainties obtained with the 3 different telescopes.
The Keplerian orbit with a linear trend has parameters of period $P=1455^{+13}
    _{-12}$ days, velocity semiamplitude $K_1=92.2^{+4.8}_{-4.7}$ m\,s$^{-1}$,
    eccentricity $e=0.533^{+0.043}_{-0.047}$ and linear velocity trend
    $\dot{\gamma}=-22.4\pm2.2$ m\,s$^{-1}$\,yr$^{-1}$.
Adopting the stellar mass of $M_\star=2.4\pm0.2\,M_\odot$ given in table
    \ref{tbl:star}, we obtain $m_2\sin{i}=7.8\pm0.7\,M_{\rm J}$ and
    $a=3.4\pm0.1$ AU for the companion.
The R.M.S. scatter of the residuals to the Keplerian fit is 12.7 m\,s$^{-1}$.
The Keplerian orbit without linear trend model yields the parameters of
    $P=2850^{+430}_{-290}$ days, $K_1=100.1^{+4.9}_{-4.8}$ m\,s$^{-1}$, and
    $e=0.697^{+0.045}_{-0.051}$, and the corresponding parameters of the
    companion are $m_2\sin{i}=9.0\pm0.9\,M_{\rm J}$ and $a=5.3^{+0.6}_{-0.4}$
    AU.
The R.M.S. scatter of the residuals of this model is 14.3\,m\,s$^{-1}$, and
    these solutions are also shown in figure \ref{fig-rv}.
Since the R.M.S scatters of residuals in two models are close to each other and
    both are comparable to the radial velocity jitter ($\sim$8\,m\,s$^{-1}$) due
    to the stellar oscillations estimated using the scaling relations of
    \citet{Kjeldsen1995},
it is difficult to determine whether the linear velocity trend should be included.
We calculated the Bayesian information criterion (${\rm BIC}=-2\ln{{\cal L}_{\rm
    max}}+k\ln{N}$; \citealt{Schwarz1978, Liddle2004, Liddle2007}) of the two
    models, where ${\cal L}_{\rm max}$ is the maximum likelihood of each model,
    $k$ denotes the number of free parameters in a model, and $N$ is the number
    of data points.
The Keplerian orbit with a linear velocity trend has smaller BIC value than the
    model without trend, but the former solution is largely constrained by the
    only 3 measurements with Subaru, and hence severely affected by the velocity
    offsets between data sets with different instruments.
The velocity trend correspond to a possible outer companion with

    $$\frac{m_{\rm c}\sin{i_{\rm c}}}{a_{\rm c}^2}\sim
    \frac{\dot{\gamma}}{G}=(0.13\pm0.01)\,M_{\rm J}\,{\rm AU}^{-2}$$

using the order-of-magnitude relation of \citet{Winn2009}, where $m_{\rm c}$,
    $i_{\rm c}$ and $a_{\rm c}$ are mass, orbital inclination, and semi-major
    axis of the companion, respectively.
This means if the companion lies outside $\sim$ 10 AU, it shall not be an
    exoplanet with $m_{\rm c}<M_{\rm b}$, where $M_{\rm b}\simeq13\,M_{\rm J}$
    is the lower mass limit of a typical brown dwarf.
On the other hand, there is no evidence that HD\,14067 has any companion star
    so far.
After removal of the orbital fitting, there is no significant peak with FAP$>$
    0.1 on the generalized Lomb-Scargle periodograms of the velocity residuals
    (figure \ref{fig-gls-res}).

\begin{figure}
    \begin{center}
    \FigureFile(16cm,){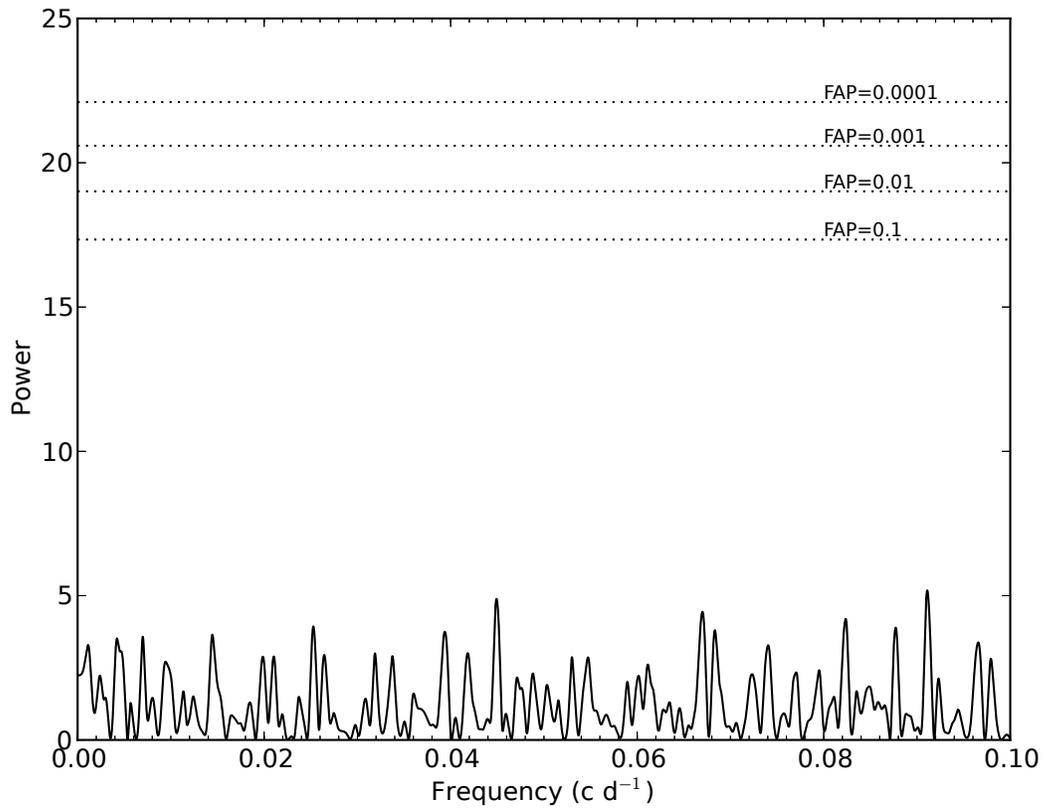}
    \end{center}
    \caption{
        Generalized Lomb-Scargle periodogram of the Hipparcos photometric data
        of HD\,14067.
        The dash horizontal lines indicate the FAP levels of $10^{-1}$,
        $10^{-2}$, $10^{-3}$, and $10^{-4}$, respectively, from the bottom to
        the top.
    }\label{fig-gls-photo}
\end{figure}

The Hipparcos satellite made a total number of 86 photometric observations for
    HD\,14067 during the period from Jan. 1990 to Feb. 1993.
The scatter of H$_{\rm p}$ magnitude is down to 0.007\,mag, but the photometry
    is not contemporaneous with the radial velocity measurements and also the
    time span ($\sim$1100\,days) is not long enough to cover a whole radial
    velocity period.
Figure \ref{fig-gls-photo} shows the generalized Lomb-Scargle periodogram of the
    HD\,14067 photometric data.
We did not find any clue that the radial velocity variation correlates with the
    brightness changes due to the stellar spots.

We also performed spectral line shape analysis for the star following the method
    in \citet{Sato2007}.
Cross correlation profiles of the two stellar templates, which were extracted
    from five I$_2$-superposed spectra at phases of velocity maximum
    (JD$\sim$2455600) and minimum (JD$\sim$2455900) using the technique by
    \citet{Sato2002}, were derived for about 90 spectral segments (4--5\AA\
    width each).
Then three bisector quantities of the cross correlation profiles, BVS, BVC, and
    BVD, were calculated, which are the velocity difference between the two flux
    level of the bisector, the difference of the BVS of the upper and lower half
    of the bisector, and the average of the bisector at three flux levels,
    respectively.
We used the flux levels of 25, 50, and 75\% of each cross correlation profile to
    calculate the above three bisector quantities.
As a result, we obtained BVS$=-8.4\pm3.8$\,m\,s$^{-1}$,
    BVC$=2.7\pm1.8$\,m\,s$^{-1}$, and BVD$=-186.7\pm7.1$\,m\,s$^{-1}$
    ($\simeq2K_1$), suggesting that the observed radial velocity variations are
    not caused by distortion of the spectral lines but by parallel shifts of
    them as expected in the case of orbital motion.

\section{Conclusion}
We report the detection of a long-period substellar companion to the G9\,III
    evolved intermediate-mass ($M=2.4\,M_\odot$) star HD\,14067 from the Subaru
    and Japan-China planet search program.
The radial velocity variation of the star can be best explained by the
    gravitational perturbation of an unseen surrounding companion.
The orbit can be well fitted with two models - a Keplerian orbit with and
    without a linear velocity trend, both of which have similar velocity
    residuals, while the former model has smaller BIC value than the latter one.
The minimum mass of the companion is $m_2\sin{i}=7.8\pm0.7\,M_{\rm J}$ (with
    velocity trend) or $m_2\sin{i}=9.0\pm0.9\,M_{\rm J}$ (without trend),
    suggesting a long-period giant planet orbiting around an evolved giant star.
It is notable that HD\,14067\,b is among the planets with the largest semi-major
    axes and orbital periods around evolved intermediate-mass stars.
The possible decreasing linear velocity trend ($-22.4$\,m\,s$^{-1}$\,yr$^{-1}$)
    may suggests an additional outer companion around this star.
Unfortunately, the time span of observations is not long enough neither to
    distinguish between these two possible orbital solutions nor to confirm the
    existence of this outer companion.
Therefore, the continued radial velocity observations are essential to
    characterize the orbital properties of this system.

\begin{figure}
    \begin{center}
    \FigureFile(16cm,){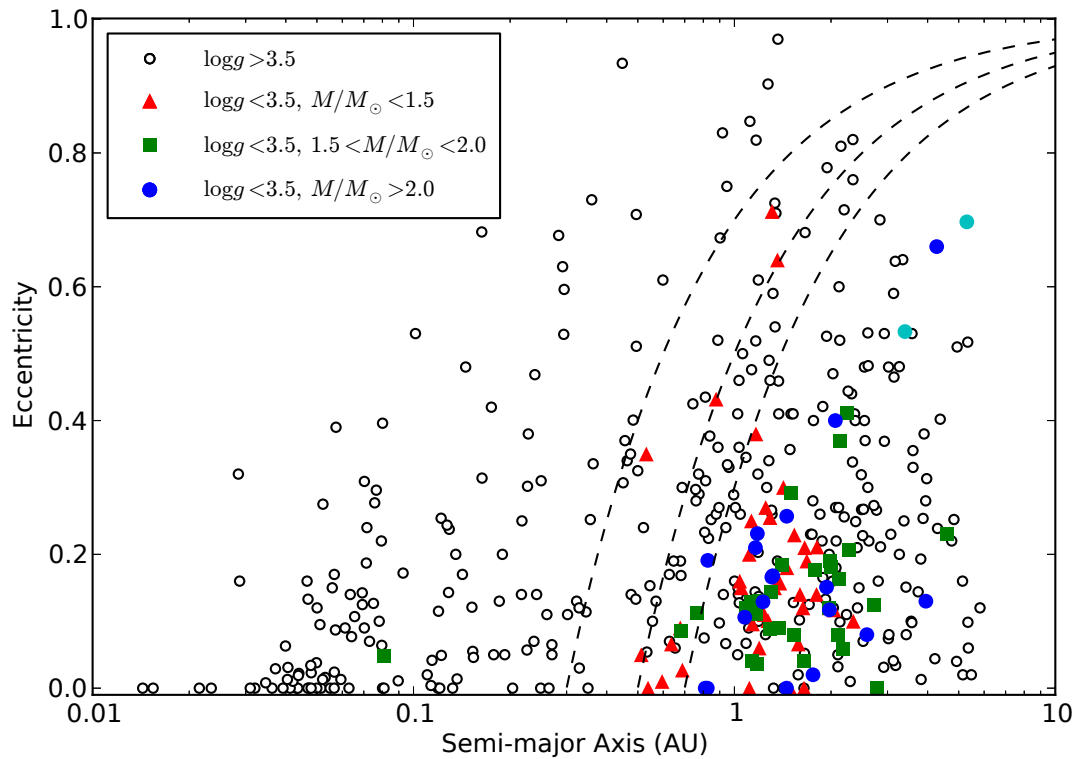}
    \end{center}
    \caption{
        Eccentricities versus semi-major axes of exoplanets discovered by radial
        velocity technique.
        Planets around low-mass ($M_\star<1.5\,M_\odot$), intermediate-mass
        ($1.5\,M_\odot<M_\star<2.0\,M_\odot$), and high-mass ($M_\star>2.0\,
        M_\odot$) evolved stars with $\log{g}_\star<3.5$ are plotted by red
        filled triangles, green filled squares, and blue filled circles,
        respectively.
        The two possible solutions of the newly detected companion HD\,14067\,b
        is plotted by the cyan circles.
        Planets around stars with $\log{g}_\star>3.5$ are plotted by open
        circles.
        Dashed lines indicate the periastron distance $[q=a(1-e)]$ of 0.3, 0.5,
        and 0.7 AU, respectively, from the left to the right.
        }
        \label{fig-ea}
\end{figure}

\begin{figure}
    \begin{center}
    \FigureFile(16cm,){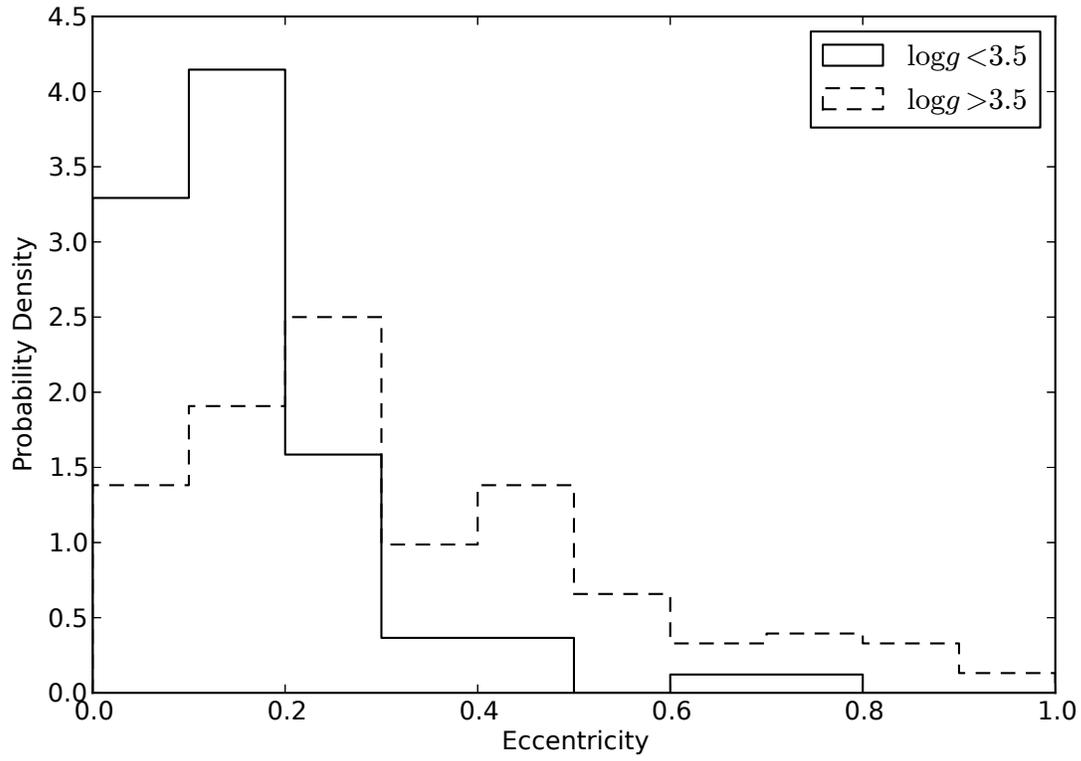}
    \end{center}
    \caption{
        Probability density distribution of eccentricities of radial velocity
        planets with 0.5 AU $<a<$ 3.0 AU around main sequence stars
        ($\log{g}_\star>3.5$), and around evolved stars ($\log{g}_\star<3.5$)
        are plotted by dashed lines and solid lines, respectively.
        }
        \label{fig-ehist}
\end{figure}

Although the fitting of Keplerian orbit with a linear trend significantly
    reduce the resulting eccentricity, orbital solutions with both models show
    the eccentricity of the HD\,14067\,b is large ($e=0.533$ with a velocity
    trend, or $e=0.697$ without velocity trend), making this planet one of the
    most eccentric ones ever discovered around stars with $M_\star>1.5\,M_\odot$.
Figure \ref{fig-ehist} shows a concentration towards low eccentricity for
    planets around off-main sequence stars ($\log{g_\star}<3.5$), while the
    eccentricity distribution of planets around main sequence stars reach its
    peak at $e=0.2\sim0.3$.
To avoid the bias due to the lack of short-period, circularized planets around
    evolved stars, only planets with semi-major axes 0.5 AU $<a<$ 3.0 AU are
    plotted.
The high eccentricity of HD\,14067\,b might originate from (a) the incomplete
    circularization due to its large distance to its host, because $|\dot{e}|$
    decreases as $a^{-8}$ \citep{Zahn1989}, or (b) the perturbation of an outer
    companion.
We plot the eccentricities against the semi-major axes of exoplanets found
    around evolved and main sequence stars with Doppler technique in figure
    \ref{fig-ea}, where the colors are coded with different ranges of stellar
    mass.
It is clearly shown that almost all planets orbiting high-mass ($M_\star>2.0\,
    M_\odot$) stars have periastron distances $[q=a(1-e)]>0.66$\,AU.
The closest blue point to HD\,14067\,b in figure \ref{fig-ea} is HD\,120084\,b,
    which also has an eccentric orbit ($a=4.3$ AU, $e=0.66$, see
    \citealt{Sato2013b}) around an evolved star with $M_*=2.39\,M_\odot$.
The large semi-major axes of these two long-period planets support the former
    possibility of the origin of the large eccentricities, which means they were
    less affected by the orbital circularization during the RGB phase compared
    with other planets around evolved stars.
We also noticed most planet-harboring evolved stars accompanied by eccentric
    planets ($e>0.4$) also exhibit linear radial velocity trends besides the
    periodic variations caused by their companions, or lie in systems with more
    than two objects.
For example, HD\,1690 (with an eccentric planet of $e=0.64\pm0.04$) shows a
    linear decreasing velocity trend of $\dot{\gamma}=-7.2\pm0.4$\,m\,s$^{-1}$
    \citep{Moutou2011}.
HD\,102272\,c ($e=0.68\pm0.06$) is the outer companion of a double-planet system
    \citep{Niedzielski2009}.
And the parent stars of HD\,137759\,b ($e=0.70\pm0.01$, \citealt{Frink2002}) and
    HD\,110014\,b ($e=0.462\pm0.069$, \citealt{deMedeiros2009}) are both members
    of double star systems.
These facts imply that the high eccentricities of such planets can be excited by
    the gravitational perturbations of additional objects, or the existences of
    these objects stop the orbital circularizations of the planets.
If the relatively small eccentricities of planets around evolved stars are
    caused by the tidal circularization when the parent stars expand during the
    RGB phase, it is expected to see higher orbital eccentricities with
    increasing semi-major axes, especially in the systems where a distant outer
    perturbator exists.
Future discoveries with longer observational baseline will be of great help to
    answer this crucial question.

\bigskip
This research is based on data collected at the Subaru Telescope, and Okayama
    Astrophysical Observatory (OAO), both operated by National Astronomical
    Observatory of Japan (NAOJ), and the 2.16m telescope at Xinglong
    Observatory, operated by National Astronomical Observatories, Chinese
    Academy of Sciences.
We are grateful to all the staff members of Subaru, OAO, and Xinglong for their
    support during the observations.
We thank students of Tokyo Institute of Technology and Kobe University for their
    kind help for the observations at Subaru and OAO.
LW is supported by the Young Researcher Grant of National Astronomical
    Observatories, Chinese Academy of Sciences.
BS was partly supported by MEXT's program "Promotion of Environmental
    Improvement for Independence of Young Researchers" under the Special
    Coordination Funds for Promoting Science and Technology and by Grant-in-Aid
    for Young Scientists (B) 20740101 from the Japan Society for the Promotion
    of Science (JSPS).
BS is supported by Grant-In-Aid for Scientific Research (C) 23540263 from JSPS
    and HI is supported by Grant-In-Aid for Scientific Research (A) 23244038
    from JSPS.
YJL and LW are supported by the National Natural Science Foundation of China
    under grants 11173031.
This research has made use of the SIMBAD database, operated at CDS, Strassbourg,
    France.

\bibliography{HD14067}

\pagebreak

\pagebreak

\pagebreak

\pagebreak

\pagebreak

\begin{figure}
    \begin{center}
    \rotatebox{-90}{
    \FigureFile(10cm,){HD14067wt.eps}
    \FigureFile(10cm,){HD14067wot.eps}}
    \end{center}
    \caption{Radial velocities of HD\,14067 observed at OAO (red circles),
        Xinglong (green triangles), and Subaru (blue squares).
        The error bar for each point includes the stellar jitter.
        The Keplerian orbits with ({\it top panel}) and without ({\it bottom
        panel}) a linear velocity trend are shown by the solid lines.        
        Residuals to the orbit fits are also shown in each panel.
        The R.M.S. is 12.7\,m\,s$^{-1}$ (with linear trend) and
        14.3\,m\,s$^{-1}$ (without linear trend), respectively.}
        \label{fig-rv}
\end{figure}

\pagebreak

\pagebreak

\pagebreak

\begin{table}
    \caption{Stellar parameters of HD\,14067}\label{tbl:star}
    \begin{center}
      \begin{tabular}{ll}
        \hline
        \hline
        Parameter                       & HD\,14067                   \\
        \hline                                                        
        Sp. type                        & G9\,III                     \\
        $\pi$ (mas)                     &  $6.12\pm0.46$              \\
        Distance (pc)                   & $163.4\pm12.3$              \\
        $V$                             & 6.53                        \\
        $B-V$                           & 1.025                       \\
        $A_V$                           & 0.14                        \\
        $M_V$                           & 0.33                        \\
        $BC$                            & $-$0.329                    \\
        $T_{\rm eff}$ (K)               &  $4815\pm100$               \\
        $\log g$                        &  $2.61\pm0.10$              \\
                                        
        [Fe/H]                          & $-0.10\pm0.08$              \\
        $v_{\rm t}$ (km\,s$^{-1}$)      &  $1.30\pm0.15$              \\
        $L\ (L_\odot)$                  &    $79\pm12$                \\
        $R\ (R_\odot)$                  &  $12.4\pm1.1$               \\
        $M\ (M_\odot)$                  &   $2.4\pm0.2$               \\
        age (Gyr)                       &  $0.69\pm0.20$              \\
        $v_{\rm macro}$ (km\,s$^{-1}$)  &  $5.44\pm0.45$              \\
        $v\sin{i}$ (km\,s$^{-1}$)       & $<1$ \citep{deMedeiros1999} \\
        $\log{A({\rm Li})}$             & 0.53 (LTE), 0.73 (NLTE)     \\
        \hline
      \end{tabular}
    \end{center}
\end{table}

\pagebreak

\begin{table}
    \caption{Orbital parameters of HD\,14067\,b}
    \label{tbl-planets}
    \begin{center}
    \begin{tabular}{lrr}
    \hline\hline
    Parameter      & \multicolumn{2}{c}{HD\,14067 b}\\
                   & with trend & without trend \\
    \hline
    $P$ (days)                                    &  1455$^{+13}_{-12}$        & 2850$^{+430}_{-290}$     \\
    $K_1$ (m\,s$^{-1}$)                           &  92.2$^{+4.8}_{-4.7}$      & 100.1$^{+4.9}_{-4.8}$    \\
    $e$                                           &  0.533$^{+0.043}_{-0.047}$ & 0.697$^{+0.045}_{-0.051}$\\
    $\omega$ (deg)                                &  109.9$^{+5.6}_{-5.7}$     & 102.1$^{+5.1}_{-5.3}$    \\
    $T_{\rm p}$ (JD$-$2,450,000)                  & 1443$^{+31}_{-37}$         & 92$^{+586}_{-862}$       \\
    $s_{\rm OAO}$ (m\,s$^{-1}$)                   & 13.8$^{+2.6}_{-2.0}$       & 14.9$^{+2.7}_{-2.1}$     \\
    $s_{\rm Xinglong}$ (m\,s$^{-1}$)              & 12.2$^{+3.0}_{-2.4}$       & 12.3$^{+3.0}_{-2.4}$     \\
    $s_{\rm Subaru}$ (m\,s$^{-1}$)                & 5.4$^{+22.2}_{-4.6}$       & 34$^{+31}_{-14}$         \\
    $\Delta$RV$_{\rm Xinglong-OAO}$ (m\,s$^{-1}$) & 19.8$^{+6.0}_{-6.0}$       & 21.5$^{+6.1}_{-6.2}$     \\
    $\Delta$RV$_{\rm Subaru-OAO}$ (m\,s$^{-1}$)   & 3.6$^{+11.3}_{-12.5}$      & $-8.7^{+24.1}_{-24.2}$   \\
    $\dot{\gamma}$ (m\,s$^{-1}$\,yr$^{-1}$)       & $-22.4^{+2.2}_{-2.2}$      & --                       \\
    $a_1\sin i$ (10$^{-3}$ AU)                    & 10.43$^{+0.50}_{-0.51}$    & 18.9$^{+1.8}_{-1.4}$     \\
    $f_1(m)$ (10$^{-7}\,M_{\odot}$)               & 0.714$^{+0.099}_{-0.91}$   & 1.10$^{+0.16}_{-0.15}$   \\
    $m_2\sin i$ ($M_{\rm J}$)                     & $7.8\pm0.7$                & $9.0\pm0.9$              \\
    $a$ (AU)                                      & $3.4\pm0.1$                & 5.3$^{+0.6}_{-0.4}$      \\
    $N_{\rm OAO}$                                 & 27                                                    \\
    $N_{\rm Xinglong}$                            & 22                                                    \\
    $N_{\rm Subaru}$                              &  3                                                    \\
    RMS (m\,s$^{-1}$)                             & 12.7                       & 14.3                     \\
    BIC                                           & 447.07                     & 458.24                   \\
    \hline
    \end{tabular}
    \end{center}
\end{table}

\pagebreak

\begin{center}
\begin{longtable}{cccc}
\caption{Radial Velocities of HD\,14067}\label{tbl-rv}\\
    \hline\hline
    JD & Radial Velocity & Uncertainty & Observatory\\
    ($-$2450000) & (m\,s$^{-1}$) & (m\,s$^{-1}$)\\
    \hline
    \endhead
    4758.11638 &     33.7 &  3.2 & OAO     \\
    4796.20144 &     51.3 &  3.3 & OAO     \\
    4820.01616 &     41.8 &  3.0 & OAO     \\
    5164.16162 &    104.3 &  3.5 & OAO     \\
    5444.14944 &    106.1 &  3.2 & OAO     \\
    5525.16330 &    113.6 &  3.2 & OAO     \\
    5545.11627 &    106.2 &  4.0 & OAO     \\
    5580.00227 &    100.2 &  5.7 & OAO     \\
    5612.94961 &    116.7 &  4.5 & OAO     \\
    5628.91329 &     87.9 &  3.8 & OAO     \\
    5786.27403 &   $-$3.9 &  3.5 & OAO     \\
    5811.19593 &  $-$15.1 &  3.9 & OAO     \\
    5853.23042 &  $-$94.8 &  5.1 & OAO     \\
    5854.22657 &  $-$88.9 &  3.5 & OAO     \\
    5879.18379 & $-$109.5 &  3.8 & OAO     \\
    5922.00916 &  $-$81.2 &  3.8 & OAO     \\
    5938.00117 &  $-$76.6 &  3.8 & OAO     \\
    5976.91135 &  $-$79.0 &  3.5 & OAO     \\
    6139.23229 &  $-$44.8 &  4.5 & OAO     \\
    6157.28469 &  $-$37.3 &  3.8 & OAO     \\
    6212.29616 &  $-$53.1 &  3.8 & OAO     \\
    6235.19638 &  $-$63.7 &  4.4 & OAO     \\
    6250.20981 &  $-$25.0 &  3.6 & OAO     \\
    6284.05633 &  $-$45.7 &  4.0 & OAO     \\
    6552.26042 &  $-$15.2 &  4.1 & OAO     \\
    6617.07400 &  $-$22.2 &  3.7 & OAO     \\
    6666.87938 &      5.8 &  4.6 & OAO     \\
    \hline
    6232.11546 &  $-$38.1 &  5.3 & Xinglong\\
    6232.13878 &  $-$28.3 &  5.1 & Xinglong\\
    6286.95306 &  $-$13.5 &  8.1 & Xinglong\\
    6286.97627 &  $-$16.9 &  7.9 & Xinglong\\
    6287.00037 &  $-$20.3 &  7.4 & Xinglong\\
    6287.02420 &  $-$23.6 &  7.7 & Xinglong\\
    6317.94617 &     10.1 &  7.8 & Xinglong\\
    6317.96965 &      3.4 &  7.6 & Xinglong\\
    6317.99284 &      2.8 &  7.9 & Xinglong\\
    6318.01605 &      4.8 &  8.6 & Xinglong\\
    6581.18750 &      6.8 &  8.3 & Xinglong\\
    6581.21073 &   $-$3.7 &  6.7 & Xinglong\\
    6581.23392 &   $-$4.8 &  8.0 & Xinglong\\
    6581.25725 &  $-$10.9 &  8.2 & Xinglong\\
    6611.11370 &     14.2 &  6.9 & Xinglong\\
    6611.13689 &     21.9 &  7.4 & Xinglong\\
    6611.16007 &     34.2 &  7.4 & Xinglong\\
    6611.18325 &     39.9 &  6.4 & Xinglong\\
    6646.03568 &      8.9 & 12.1 & Xinglong\\
    6646.05898 &     10.6 & 11.6 & Xinglong\\
    6646.08226 &      7.1 & 13.3 & Xinglong\\
    6646.10544 &      5.4 & 12.5 & Xinglong\\
    \hline        
    4364.95416 &     44.2 &  4.8 & Subaru  \\ 
    4470.70735 &   $-$5.2 &  4.1 & Subaru  \\ 
    4698.90077 &     46.0 &  5.0 & Subaru  \\
   \hline
\end{longtable}
\end{center}

\end{document}